\documentclass[runningheads]{llncs}
\usepackage{floatrow}
\newfloatcommand{capbtabbox}{table}[][\FBwidth]
\usepackage{longtable}
\usepackage[T1]{fontenc}
\usepackage{hyperref}
\usepackage{float}
\usepackage{tabularx}
\usepackage{makecell}
\usepackage{multirow}
\usepackage{booktabs}
\usepackage{amsfonts} 
\usepackage{graphicx}
\usepackage{amsmath}
\usepackage{color}
\usepackage{caption}
\usepackage[misc]{ifsym}

\begin{document}
\title{A Probabilistic Hadamard U-Net for MRI Bias Field Correction}
%
%

\author{Xin Zhu$^{1,2}$$^{(\textrm{\Letter})}$, 
Hongyi Pan$^2$, 
Batuhan Gundogdu$^3$,
Debesh Jha$^2$, 
Yury Velichko$^2$, Adam B. Murphy$^2$, Ashley Ross$^2$, Baris Turkbey$^4$, Ahmet Enis Cetin$^1$ and Ulas Bagci$^2$}
\authorrunning{Xin Zhu et al.}
%
\institute{$^1$Department of ECE, University of Illinois Chicago, Chicago, USA\\ 
$^2$Machine and Hybrid Imaging Lab, Northwestern University, Chicago, USA\\
$^3$Department of Radiology, University of
Chicago, Chicago, USA\\
$^4$Molecular Imaging Branch, NCI, National Institutes of Health, Bethesda, MD, USA\\
}

\maketitle              
\begin{abstract}
Magnetic field inhomogeneity correction remains a challenging task in MRI analysis. Most established techniques are designed for brain MRI by supposing that image intensities in the identical tissue follow a uniform distribution. Such an assumption cannot be easily applied to other organs, especially those that are small in size and heterogeneous in texture (large variations in intensity) such as the prostate. To address this problem, this paper proposes a probabilistic Hadamard U-Net (PHU-Net) for prostate MRI bias field correction. First, a novel Hadamard U-Net (HU-Net) is introduced to extract the low-frequency scalar field. HU-Net converts the input image from the time domain into the frequency domain via Hadamard transform. In the frequency domain, high-frequency components are eliminated using the trainable filter (scaling layer), hard-thresholding layer, and sparsity penalty. Next, a conditional variational autoencoder is used to encode possible bias field-corrected variants into a low-dimensional latent space. Random samples drawn from latent space are then incorporated with a prototypical corrected image to generate multiple plausible images. Experimental results demonstrate the effectiveness of PHU-Net in correcting bias-field in prostate MRI with a fast inference speed. It has also been shown that prostate MRI segmentation accuracy improves with the high-quality corrected images from PHU-Net. The codes are available at \href{https://github.com/Holmes696/Probabilistic-Hadamard-U-Net}{https://github.com/Holmes696/Probabilistic-Hadamard-U-Net}.

\footnote{Corresponding auhtor's email: \email{xzhu61@uic.edu}. This work is supported by the NIH funding: R01-CA246704, R01-CA240639, U01-DK127384-02S1, and U01-CA268808, and NSF IDEAL 2217023.}

\end{abstract}

\section{Introduction}
\label{sec: Introduction}

The main magnetic field in an Magnetic Resonance Imaging (MRI) scanner is not perfectly uniform across the imaging volume, making the quantification tasks (detection, diagnosis, segmentation) challenging and sub-optimal. To address this non-uniformity problem in MRI, one critical step is to correct bias field ~\cite{van1999automated,juntu2005bias}. The bias field refers to a low-frequency multiplicative field arising from various sources, such as coil sensitivity variations and radio-frequency inhomogeneities. It distorts intensity distributions and impacts the accuracy of subsequent downstream tasks, most importantly image segmentation algorithms.


Numerous approaches have been proposed for bias field removal. Among them, the N4 bias field correction algorithm (N4ITK) has gained popularity for its effectiveness in rectifying inhomogeneities within brain MRI by assuming uniform intensity across distinct types of brain tissues~\cite{tustison2009n4itk,tustison2010n4itk,chen2021abcnet}. However, this condition is not valid for prostate MRI because there exist large intensity variations within the same prostate tissue. Additionally, N4ITK corrects the bias field with a high computation cost, leading to a low inference speed. To improve inference speed, convolutional layers-based frameworks~\cite{simko2022mri,sridhara2021bias} were proposed. Yet, these methods are limited by their accuracy, efficiency, and generalization ability.
Alternatively, adversarial learning~\cite{chen2021abcnet,goldfryd2021deep} has emerged as a promising avenue for addressing non-uniformities in medical imaging. This technique contains two models: while a \textit{generator} corrects bias fields, a \textit{discriminator} distinguishes corrected images from ground truth images. Although adversarial networks enhance correction accuracy, they still face challenges such as model collapse, sensitivity to hyperparameters, and longer training times. 

The most crucial step of bias field correction is to extract the low-frequency multiplicative field; therefore, proper frequency analysis can improve the quality of the corrected images. At present, orthogonal transforms including discrete Fourier transform~\cite{chi2020fast}, discrete cosine transform~\cite{badawi2021discrete} and Hadamard transform~\cite{pan2023hybrid} have been widely employed in neural networks to extract low-frequency features hidden in the transform domain. In this family, the Hadamard transform is the most implementation-efficient because its transform matrix elements take only $+1$ and $-1$ values. To the best of our knowledge, there is no literature study combining U-Nets (or other neural networks) with orthogonal transforms for MRI bias field correction.



In this study, we propose a probabilistic Hadamard U-Net (PHU-Net) to correct the bias field of MRI and we demonstrate its effectiveness in prostate MRI due to its unique challenges of large intensity variations and complexity of tissue interactions. More specifically, our contributions are summarized as follows: (1) We develop a novel Hadamard U-Net based on the trainable filters and hard-thresholding layers to obtain an efficient representation of the underlying MR scalar field. (2) We incorporate a conditional variational autoencoder (CVAE) to model the joint distribution of all intensity values in ground truth and generate extensive plausible bias field corrected images. (3) We design a hybrid loss optimization composed of Kullback Leibler divergence (KLD) loss~\cite{ng2011sparse}, total variation loss, and mean squared error (MSE) loss to promote sparsity in the learned representations, resulting in improved robustness and generalization to diverse MRI datasets. 
(4) Through comprehensive experiments on several benchmark datasets, we validate the superior efficacy of PHU-Net compared to state-of-the-art bias field correction techniques, as well as improved segmentation performance. This demonstrates the potential of PHU-Net in enhancing the quality and reliability of medical image analysis pipelines.


\begin{figure}[h]
    \centering
    \includegraphics[width=1\linewidth]{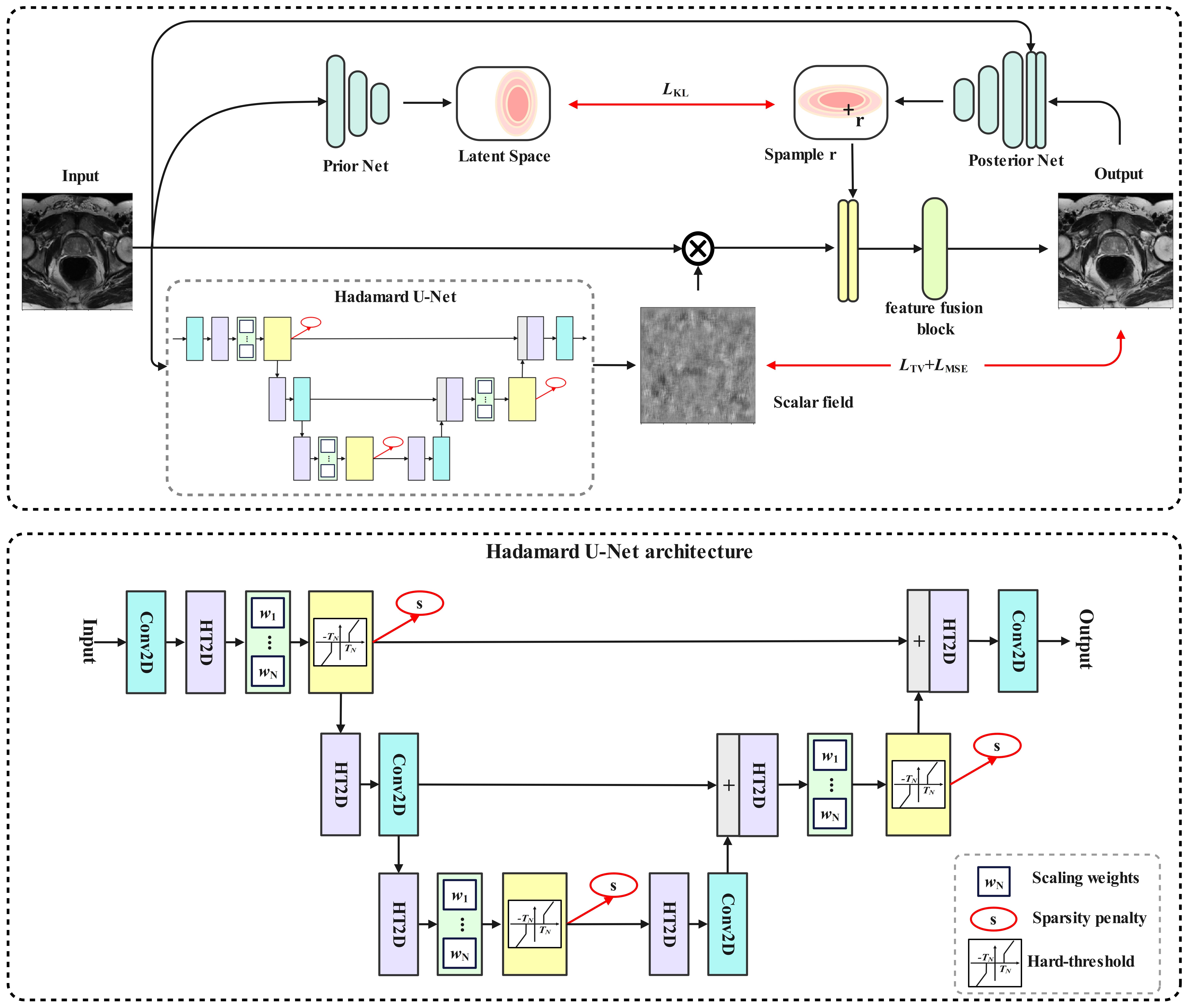}
    \caption{Probabilistic Hadamard U-Net.}
    \label{fig:Hadamard UNet}
\end{figure}
\section{Methods}
The overall framework of the PHU-Net is presented in Fig.~\ref{fig:Hadamard UNet}. Our supervised pipeline contains two inter-connected modules: a Hadamard U-Net (HU-Net) and a conditional variational autoencoder (CVAE).


\subsection{Hadamard U-Net} 
The Hadamard U-Net (HU-Net) extracts the scalar field, which is the inverse of the bias field. 
Innovatively, the Hadamard transform (HT) is applied for the frequency analysis. 
The HT of an $M\times M$ input matrix $\mathbf{X}$ is computed as:
\begin{equation}
\begin{split}
   \widehat X_{a,b} =\frac{1}{M} \sum_{c=0}^{M-1}\sum_{d=0}^{M-1}X_{c,d}(-1)^{\sum_{j=0}^{log_2M-1}a_jc_j+b_jd_j},
\end{split}
\end{equation}
where $M$ is an integer power of 2, and $a_j$, $b_j$, $c_j$ and $d_j$ are the $j$-bits in the binary representations of $a$, $b$, $c$ and $d$, respectively. 
Next, we take three steps to emphasize energy in the low-frequency band where the bias field lies: 
\textbf{First}, each entry of $ \widehat{\mathbf{X}}$ is multiplied with a trainable scaling weight: 
\begin{equation}
    \widetilde{\mathbf{X}} = \mathbf{W}\circ \widehat{\mathbf{X}},
\end{equation}
where $\circ$ denotes the element-wise product and $\mathbf{W}$ is the scaling weights. 
Applying scaling on the Hadamard coefficients is inspired by the Hadamard convolution theorem~\cite{uvsakova2002walsh}, stating that the Hadamard-domain element-wise product is similar to the time-domain dyadic convolution: 
\begin{equation}
\mathbf{m} *_d {\mathbf{n}} = \mathcal{T}^{-1}\left(\mathcal{T}(\mathbf{m})\circ\mathcal{T}(\mathbf{n})\right), 
\end{equation}
where $\mathbf{m}$ and $\mathbf{n}$ are input vectors. $\mathcal{T}(\cdot)$ denotes the HT. $*_d$ is the dyadic convolution. Therefore, the scaling layer in the HT domain works as the convolution filter in the time-domain. On the other hand, the scaling layer requires significantly less computational overhead than the time-domain convolutional filtering.

\textbf{Second}, a trainable hard-thresholding layer $\mathcal{H}_\mathbf{T}(\cdot)$ diminishes the influence of the high-frequency band: 
\begin{equation}
    \mathbf{Z} = \mathcal{H}_\mathbf{T}(\widetilde{\mathbf{X}})=\mathcal{C}_\mathbf{T}(\widetilde{\mathbf{X}})+\text{sign}(\mathcal{C}_\mathbf{T}(\widetilde{\mathbf{X}}))\circ \mathbf{T},\label{eq: HST}
\end{equation}
\begin{equation}
    \mathcal{C}_\mathbf{T}(\widetilde{\mathbf{X}}) = \text{sign}(\widetilde{\mathbf{X}})\circ \text{ReLU}(|\widetilde{\mathbf{X}}|-\mathbf{T}),\label{eq: ST}
\end{equation}
where $\text{ReLU}(\cdot)$ stands for the standard ReLU activation function~\cite{fukushima1969visual} and $\mathbf{T}\in\mathbb{R}^{M\times M}$ is non-negative trainable threshold parameters determined using the back-propagation algorithm. Specially, Eq. (\ref{eq: ST}) can be simplified as:
\begin{equation}
    \mathbf{Z} = \mathcal{C}_\mathbf{T}(\widetilde{\mathbf{X}}) = \text{copysign}(|\widetilde{\mathbf{X}}|-\mathbf{T}, \widetilde{\mathbf{X}}),
\end{equation}
where $\text{copysign}(\cdot, \cdot)$ replaces the first argument's sign bit with the second argument's sign bit. In this implementation, no multiplication is required. The hard-thresholding layer can constrain the high-frequency band by shifting the small entries to $0s$. In the frequency domain, the low-frequency magnitudes are usually much larger than the high-frequency magnitudes. Hence, such a shifting with a proper threshold can remove the high-frequency components while preserving the low-frequency components. 

\textbf{Third}, the output of the hard-thresholding layer is penalized using the Kullback Leibler divergence (KLD)-based sparsity constraint.  After HU-Net, the original input is element-wise multiplied by the scalar field to achieve the prototypical bias field corrected image. 

\subsection{Conditional Variational Autoencoder}
Since the intensities in the prostate MRI data follow a non-uniform distribution, the conditional variational autoencoder (CVAE)~\cite{kohl2018probabilistic} is applied to establish the joint probability of all intensities in the ground truth $\mathbf{Y}$. 
As shown in Fig.~\ref{fig:Hadamard UNet}, the \textit{prior network} maps the raw image $\mathbf X$ to a Gaussian distribution parameterized by its mean $\mathbf{u}\in\mathbb{R}^{D}$ and covariance $\mathbf{V}\in\mathbb{R}^{D\times D}$ in the latent space, where $D$ is the dimension of the latent space.
Next, the raw image and the ground truth are concatenated together as the input of the \textit{posterior network}. This posterior network predicts the distribution of bias field-corrected images conditioned on a raw image. In the training step, 
the random samples $\mathbf{r}_i\in\mathbb{R}^{D}$ are drawn from the posterior distribution $F$:
\begin{equation}
    \mathbf{r}_q\sim F(\cdot\mid \mathbf X,\mathbf Y)=\mathcal{N}(\mathbf{u}(\mathbf X, \mathbf Y;q),\mathbf{V}(\mathbf X, \mathbf Y; q)),
\end{equation}
where $q$ represents the weights of the posterior network. In the testing stage, random samples are drawn from the prior distribution $G$:
\begin{equation}
    \mathbf{r}_p\sim G(\cdot\mid \mathbf X)=\mathcal{N}(\mathbf{u}(\mathbf X;p),\mathbf{V}(\mathbf X; p)),
\end{equation}
where $p$ stands for the prior network weights. Then, the random samples are incorporated with HU-Net output. 
Next, the feature fusion block filters all information to obtain the corrected MRI data. Furthermore, a KLD loss function imposes a penalty on differences between the outputs of the prior and posterior networks.

\subsection{Loss Function}
The overall loss function is composed of three parts: KLD loss $\mathcal{L_{KL}}$, total variation loss $\mathcal{L_{TV}}$~\cite{osher2005iterative} and mean squared error (MSE) loss $\mathcal{L_{MS}}$:
\begin{equation}
    \mathcal{L_C}= \lambda_1\mathcal{L_{KL}}({F}||{G})+\lambda_2\sum_{i=1}^{L}\mathcal{L_{KL}}(\delta({\mathbf{Z}}_i)||\beta)+\lambda_3\mathcal{L_{TV}}(\mathbf{U})+\lambda_4\mathcal{L_{MS}}(\mathbf{I},\mathbf{O}),\label{eq:loss}
\end{equation}
where ${F}$ and ${G}$ represent the posterior distribution and the prior distribution respectively as shown in Fig.~\ref{fig:Hadamard UNet}. $\delta(\cdot)$ stands for the sigmoid function. ${\mathbf{Z}}_i$ is the average output of $i$-th hard-thresholding layer. $L$ is the number of sparsity constraints applied. $\beta$ denotes a sparse parameter.  $\mathbf{U}$ is the output of Hadamard U-Net. $\mathbf{I}$ and $\mathbf{O}$ represent the input and output respectively. $\lambda_1$, $\lambda_2$, $\lambda_3$ and $\lambda_4$ are the constant weights. 

We impose sparsity on ${\mathbf{Z}}_i$ by minimizing the term $\mathcal{L_{KL}}(\delta(\mathbf{Z}_i)||\beta)$.
Additionally, variance between $F$ and $G$ is penalized using $\mathcal{L_{KL}}(F||G)$. Moreover, we employ $\mathcal{L_{TV}}(\mathbf{U})$ to keep spatial gradients of $\mathbf{U}$ sparse. Furthermore, the difference between output and ground truth is minimized by decreasing $\mathcal{L_{MS}}(\mathbf{I},\mathbf{O})$.

\section{Experiments}

\subsection{Datasets and Implementation}

\subsubsection{Datasets} \label{sec: Datasets and Implementation}
We use four distinct publicly available MRI prostate datasets (all T2-weighted) in our experiments: HK dataset~\cite{litjens2014evaluation}, UCL dataset~\cite{litjens2014evaluation}, HCRUDB dataset~\cite{lemaitre2015computer} and AWS dataset~\cite{antonelli2022medical}. 
Since N4ITK is one of the most prevalent MRI bias field correction methods in the last decades, we employ it to clean all scans in these datasets and utilize processed scans as a reference baseline that serves as a substitute for ground truth.
To validate the generalization performance of different methods, all models are trained on the UCL dataset and then tested on the HK, HCRUDB, and AWS datasets. In summary, the training set contains 318 scans, while the three testing sets contain 288, 1,216, and 904 scans, respectively. All these scans are reshaped into $256\times256$ in-plane resolution while they have different sizes of slices.


\subsubsection{Implementation details}
The HU-Net includes 4 convolutional layers. The kernel sizes are $16\times16$, $7\times7$, $7\times7$ and $16\times16$, respectively.
In CVAE, the prior and the posterior nets share a similar encoder structure. This encoder structure is comprised of two $3\times3$ convolutional blocks with
32 and 64 filters. Each convolutional block consists of 4 convolutional layers activated by the ReLU function. 
After the first block, a 2$\times$2 average pooling layer with stride 2 downsamples the image tensors. 
Moreover, the feature fusion block shown in Fig.~\ref{fig:Hadamard UNet} contains three 1$\times$1 convolutional layers. Each of the first 2 convolutional layers contains 32 filters, and they are activated by the ReLU function. The third convolutional layer contains 1 filter. Our model is trained using the AdamW optimizer~\cite{loshchilov2017decoupled} with a batch size of 128 and a learning rate of 0.0001 under 100 training epochs. $\lambda_1$, $\lambda_2$, $\lambda_3$ and $\lambda_4$ in Eq.~(\ref{eq:loss}) are set to 10, 0.1, 1, and 1, respectively. 

\subsubsection{Performance metrics} Models are evaluated on the following metrics: Coefficient of variation (CV)~\cite{sled1998nonparametric,likar2001retrospective,madabhushi2005interplay}, signal-to-noise ratio (SNR)~\cite{welvaert2013definition} to evaluate the bias field correction performance, and Dice~\cite{laradji2021weakly}, positive
predictive value (PPV)~\cite{wang2020chexlocnet}, and intersection over union (IoU)~\cite{jiang2021aiu} to assess prostate segmentation performance. 
In general, a lower CV and higher other metrics indicate a better model performance.

           

\begin{table}[t]
\caption{Comparison experiments on CV and SNR.}
\label{Tab: Comparison experiments}
\begin{tabularx}{\textwidth}{ X| X X X |X} \hline 
{{\textbf{Metrics}}}&{}&\makecell{\textbf{CV}$\downarrow$}&{}&\makecell{\textbf{SNR}$\uparrow$}  \\ \hline
        {{\textbf{Dataset}}} &\makecell{\textbf{HK}}&\makecell{\textbf{AWS}}&\makecell{\textbf{HCRUDB}} &\makecell{\textbf{HK}} \\ 
\hline
            Original     &\makecell{78.24}    &\makecell{59.13}    &\makecell{53.78} &\makecell{24.64} \\
            N4ITK            &\makecell{78.24}    &\makecell{56.76}    &\makecell{52.41} &\makecell{24.64}\\
            CAE &\makecell{66.97}    &\makecell{62.51}    &\makecell{51.39} &\makecell{13.18}  \\
            ITCNN  &\makecell{78.25}    &\makecell{59.02}    &\makecell{53.69} &\makecell{23.50} \\    
           \textbf{PHU-Net} &\makecell{\textbf{65.95}}    &\makecell{\textbf{54.14}} &\makecell{\textbf{43.39}} &\makecell{\textbf{27.05}} \\
\hline     
\end{tabularx}
\end{table}

\begin{figure}[t]
    \centering
    \includegraphics[width=1\linewidth]{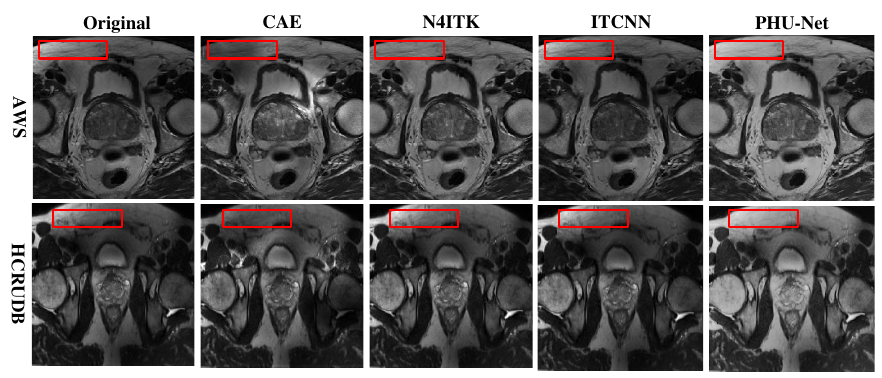}
    \caption{Comparison between the original and the corrected MRIs on the AWS and HCRUDB dataset. Last column is our proposed method.
}
    \label{fig: compare}
\end{figure}
\vspace{-2mm}
\subsection{Experimental Results}
\vspace{-1mm}
We compare the proposed PHU-Net with three up-to-date bias field correction methods, including N4ITK~\cite{tustison2010n4itk}, convolutional autoencoder (CAE)~\cite{sridhara2021bias} and implicitly trained CNN (ITCNN)~\cite{simko2022mri}.

Table \ref{Tab: Comparison experiments} compares the CV results of our proposed PHU-Net versus other state-of-the-art methods on three test datasets. Compared with N4ITK, PHU-Net reduces CV from 78.24 to 65.95 (15.71\%), from 56.76 to 54.14 (4.62\%), and from 52.41 to 43.39 (17.21\%) on the HK, AWS and HCRUDB datasets, respectively. PHU-Net also outperforms CAE and ITCNN, indicating that PHU-Net has a better capability for removing intensity variations regardless of scanner types.
Additionally, Table \ref{Tab: Comparison experiments} also presents SNR values for the MRI dataset corrected by different methods. In this comparison, models are only evaluated on the HK dataset because other datasets do not contain a clear background as the HK dataset, essential for SNR calculation. As a result, PHU-Net achieves the highest SNR. Furthermore, only PHU-Net improves SNR compared to the original images, which demonstrates our proposed method introduces no additional noise to the region of interest during the bias field correction stage. 

Experimental results are visualized in Fig.~\ref{fig: compare}. 
Images processed by PHU-Net exhibit superior intensity uniformity compared to those processed by other baseline methods. 


\begin{table}[t]
\caption{Ablation experiments on CV and SNR.}
\label{Tab: Ablation experiments}
\begin{tabularx}{\textwidth}{ X| X X X |X} \hline 
{{\textbf{Metrics}}}&{}&\makecell{\textbf{CV}$\downarrow$}&{}&\makecell{\textbf{SNR}$\uparrow$}  \\ \hline
        {{\textbf{Dataset}}} &\makecell{\textbf{HK}}&\makecell{\textbf{AWS}}&\makecell{\textbf{HCRUDB}} &\makecell{\textbf{HK}} \\ 
\hline
            No scaling     & \makecell{67.38}   &\makecell{54.92}    &\makecell{\textbf{43.12}}& \makecell{25.69} \\
            No sparsity    &\makecell{71.36}    &\makecell{58.60}    &\makecell{49.19} &\makecell{25.76} \\
            No threshold &\makecell{67.44}    &\makecell{54.83}    & \makecell{43.13}  &\makecell{25.43}\\
            No TV loss  & \makecell{67.21}   &\makecell{55.07}    &\makecell{43.77} &\makecell{25.73}  \\     
            No CVAE     &\makecell{68.27}    &\makecell{56.15}  &\makecell{44.09} &\makecell{26.25}\\
            \textbf{PHU-Net} &\makecell{\textbf{65.95}} &\makecell{\textbf{54.14}}&\makecell{{43.39}} &\makecell{\textbf{27.05}} \\
\hline     
\end{tabularx}
\end{table}
\vspace{-4mm}

\subsubsection{Ablation study}
To verify each module in PHU-Net, ablation studies are carried out in Table~\ref{Tab: Ablation experiments}. When the scaling layer, sparsity, hard-thresholding layer, TV loss, or CVAE are removed, SNR is reduced by 2.96\% at least and 5.99\% at most. Although CV improves slightly in the HCRUDB dataset when the scaling layer is not present, it degrades significantly in the HK and AWS datasets. When there is no CVAE, PHU-Net only contains HU-Net. Its results are still better than those of other models in Table \ref{Tab: Comparison experiments}. 

\begin{table}[t]
\caption{Segmentation experiment on the HK dataset.}
\label{Tab: seg}
\begin{tabularx}{\textwidth}{ X X X X} \hline 
{\textbf{Metrics}}&\makecell{\textbf{Dice(\%)}}&\makecell{\textbf{IoU(\%)}}&\makecell{\textbf{PPV(\%)}}  \\ 
\hline
            Original &\makecell{71.43}    &\makecell{66.22} &\makecell{76.22}  \\
            N4ITK &\makecell{72.38} & \makecell{67.38} &\makecell{75.74} \\
            CAE&\makecell{68.20} & \makecell{63.04} &\makecell{74.07}  \\
            ITCNN&\makecell{75.51} & \makecell{70.27}  &\makecell{80.06}\\    
            \textbf{PHU-Net}  &\makecell{\textbf{76.86}} & \makecell{\textbf{71.72}}   &\makecell{\textbf{82.34}}\\   
\hline     
\end{tabularx}
\end{table}
\vspace{-3mm}
\subsubsection{Segmentation experiment}
Corrected images are further utilized in the segmentation experiments. Here, a traditional probabilistic U-Net~\cite{kohl2018probabilistic} is used as the segmentation model, containing a standard U-Net and a CVAE. The top 50\% of the corrected dataset is utilized as the training dataset while the remaining is employed as the testing dataset. The MSE loss function is applied. Other implementation details are the same as in \cite{kohl2018probabilistic}. As shown in Table \ref{Tab: seg}, it achieves the highest Dice, IoU, and PPV on the images corrected by PHU-Net. Therefore, PHU-Net ensures more accurate and reliable segmentation of the underlying tissue structures than other correction models.
\vspace{-2mm}
\subsubsection{Runtime analyses} Another benefit of PHU-Net lies in its inference efficiency. For one single NIfTI image file (~\~15 slices), the bias field correction runtime of N4ITK, CAE, ITCNN, and PHU-Net are around 44.26, 1.06, 1.23, and 1.13 seconds, respectively. These tests are executed on a computer with an Intel Core i7-12700H CPU. Compared to N4ITK, our proposed model exhibits nearly 40 times faster processing speed, and similar inference times with ITCNN and CAE while better in performance than them.


\vspace{-3mm}
\section{Conclusion}
\vspace{-3mm}
In this paper, we developed a novel framework named PHU-Net for  MRI bias field correction.  The key idea of PHU-Net is to combine a U-Net structure with transform domain methods, trainable filters, and hard-thresholding layers to extract biased field information.  Experimental results showed that PHU-Net achieved a better bias field correction performance than other state-of-the-art methods with a fast running speed. What is more, PHU-Net significantly improved the segmentation accuracy of the probabilistic U-Net.

%
%
%
%
{
\small
\bibliographystyle{splncs03}
\bibliography{refs}
}

\end{document}